\documentclass[twocolumn,prl]{revtex4}

\usepackage[english]{babel}
\usepackage{amssymb}
\usepackage{amsfonts}
\usepackage{amsmath}

\usepackage{graphicx}

\begin{document}

\title{Insight into Resonant Activation in Discrete Systems}

\author{O. Flomenbom and J. Klafter}
\address{\it School of Chemistry, Raymond \& Beverly Sackler Faculty Of Exact Sciences,\\  Tel-Aviv University, \it
Tel-Aviv 69978, Israel}%

\thanks{We thank Attila Szabo for fruitful discussions.}%


\date{\today}

\begin{abstract}
The resonant activation phenomenon (RAP) in a discrete system is
studied using the master equation formalism. We show that the RAP
corresponds to a non-monotonic behavior of the frequency dependent
first passage time probability density function ({\it pdf}). An
analytical expression for the resonant frequency is introduced,
which, together with numerical results, helps understand the RAP
behavior in the space spanned by the transition rates for the case
of reflecting and absorbing boundary conditions. The limited range
of system parameters for which the RAP occurs is discussed. We
show that a minimum and a maximum in the mean first passage time
(MFPT) can be obtained when both boundaries are absorbing.
Relationships to some biological systems are suggested.
\end{abstract}
\maketitle

\begin{center}
{\bf {1. INTRODUCTION}}\end{center}

Noise induced escape of a particle from a potential well has been
a fundamental way for describing various processes in biology,
chemistry, and physics, since the seminal work of Kramers
\cite{1}. More recently, it has been suggested that for some
systems the potential itself fluctuates in time. A few examples
are: the transport of ions and bio-polymers through membrane
channels \cite{2}-\cite{4}, enzymatic kinetics \cite{5}, and the
rebinding of ligand-protein complexes \cite{6,7}. The basic
formulation of noise induced escape from a fluctuating environment
is obtained by making the potential term of the stochastic
differential equation (e.g. a white noise overdamped Langevin
equation) change with a frequency $\gamma$ between two states.
Doering and Gadoua showed that for a fluctuating system, the MFPT,
$\tau$, from a reflecting boundary to an absorbing boundary, may
show a minimum as a function of $\gamma$ \cite{8}. The occurrence
of a minimum in $\tau(\gamma)$ was termed the resonant activation
phenomenon (RAP). This has been followed by an extensive
theoretical work to understand the nature of the RAP
\cite{9}-\cite{16}, along with experimental efforts to find
systems displaying RAP \cite{2,17}. The theoretical works have
been mainly focused on checking the effect of different potentials
on the RAP.

In this paper we study the discrete case RAP using coupled master
equations (ME). We show that the RAP is only one of the properties
that stem from the non-monotonic behavior of the frequency
dependent first passage times (FPT) {\it pdf}, $F_{\gamma}(t)$,
and which are related to frequency dependent minima in the first
and higher moments of $F_{\gamma}(t)$. We introduce an analytical
expression for the dependence of the frequency that minimizes
$\tau(\gamma)$ on the system transitions rates. We show that the
RAP is obtained only when certain conditions imposed on the
transition rates are fulfilled. Analyzing these conditions we come
up with an instructive understanding regarding the nature of the
RAP. In addition, a novel behavior of the MFPT is obtained when
changing the reflecting boundary into an absorbing one: the
coexistence of a minimum and a maximum in $\tau(\gamma)$.
\begin{figure}[b]
\includegraphics[width=0.5 \linewidth,angle=0]{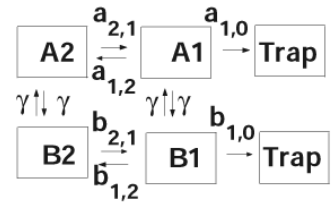}
\caption{Schematic illustration of the exit problem in a
fluctuating environment for $n=2$. For an invariant system
$a_{1,0}$=$a_{2,1}$=$a_f$, $a_{1,2}$=$a_b$, and
$b_{1,0}$=$b_{2,1}$=$b_f$, $b_{1,2}$=$b_b$. For the particular
invariant birth-death system $a_{b}$=0, and $b_{f}$=0.}
\end{figure}

\begin{center}
{\bf {2. THEORETICAL FRAMEWORK}}\end{center}

We describe the escape from a system that fluctuates between two
configurations $A$ and $B$, using the coupled ME:
\begin{equation}\label{1}
   \frac{\partial}{\partial t}
       \begin{array}{cc}
       \left( \begin{array}{c}
       \overrightarrow{P}_A(t) \\
       \overrightarrow{ P}_B(t) \\
       \end{array}\right)
        =
       \left( \begin{array}{cc}
       \mathbf{A}-\mathbf{I}\gamma & \mathbf{I}\gamma \\
       \mathbf{I}\gamma & \mathbf{B}-\mathbf{I}\gamma \\
       \end{array}\right)
       \left(\begin{array}{c}
       \overrightarrow{P}_A(t) \\
       \overrightarrow{P}_B(t) \\
       \end{array}\right).
  \end{array}
 \end{equation}
$\overrightarrow{P}_A(t)$ ($\overrightarrow{P}_B(t)$) is an
$n$-dimensional column vector, whose $j$ element is the {\it pdf}
to occupy site $j$ of the {\it A} ({\it B}) configuration at time
$t$. The transition between each site $j$ in one configuration and
its counterpart in the second configuration occurs with a flipping
frequency $\gamma$, see Fig. 1. $\mathbf{I}$ is the unit matrix of
$n$ dimensions introduced in Eq. (1) to indicate the
configurational coupling. Movement along each of the
configurations, $A$ and $B$, is governed by the square
$n$-dimensional tridiagonal propagation matrices $\mathbf{A}$ and
$\mathbf{B}$, respectively, whose elements are the transition
rates (Fig. 1). The choice of the matrices $\mathbf{A}$ and
$\mathbf{B}$ corresponds to an equivalent choice of potential
profiles and boundary conditions in the continuum case. In what
follows we set a reflecting boundary at site $j$=$n$, and an
absorbing boundary, as a trap, at site $j$=$0$, unless otherwise
indicated. Fig. 1 shows a schematic illustration of the coupled
system for $n=2$.

The FPT {\it pdf} is defined by
$F_{\gamma}(t)=\partial(1-S_{\gamma}(t))/\partial t$, where
$S_{\gamma}(t)$ is the survival probability; namely, the
probability of not reaching the site $j$=$0$ until time $t$.
$S_{\gamma}(t)$ is obtained by summing the elements of the vector
that solves Eq. (1),
$S_{\gamma}(t)=\overrightarrow{U}_{2n}\mathbf{E}e^{\mathbf{D}t}\mathbf{E}^{-1}\overrightarrow{P}_{2n}(0)$.
Here $\overrightarrow{U}_{2n}$ is the summation row vector of $2n$
dimensions, $\overrightarrow{P}_{2n}(0)$ is the initial condition
column
vector,~$[\overrightarrow{P}_{2n}(0)]_j=(\delta_{x,j}P_{A,0}+\delta_{x+n,j}P_{B,0})$,
where $x$ is the initial site, and the process starts in the $A$
($B$) configuration with probability $P_{A,0}$ ($P_{B,0}$). Unless
otherwise specified, we use $x$=$n$ as a starting site, and
$P_{A,0}$=$P_{B,0}$=$1/2$, as suggested from the single
configurational flipping frequency. The definite negative real
part eigenvalues matrix $\mathbf{D}$ is obtained through the
similarity transformation:
$\mathbf{D}=\mathbf{E}^{-1}\mathbf{H}\mathbf{E}$, where
$\mathbf{H}$ is the matrix given on the right hand side of Eq.
(1), and $\mathbf{E}$ and $\mathbf{E}^{-1}$ are the eigenvectors
matrix, and its inverse, of $\mathbf{H}$.

\begin{center}
{\bf {3. RESULTS AND DISCUSSION}}\end{center}

We start by computing $F_{\gamma}(t)$ for an invariant birth-death
system. By an invariant system we mean that the transition rates
are independent of the site index, $j$; namely,
$m_{j,j-/+1}$=$m_{f/b}$ for configuration $M$, where $m_{f/b}$
represents $a_{f/b}$ and $b_{f/b}$ transition rates, and $M$
stands for $A$ and $B$. A birth-death system means that the
movement in each configuration occurs only in one direction, i.e.
$a_b$=$b_f$=$0$. Clearly, the term birth-death indicates that the
particle [when simulating Eq. (1)] can move only towards its
"death" (the trap) when it is subjected to the dynamics of the $A$
configuration, and in this sense, when flipping to the $B$
configuration occurs it is "born" (or "resurrected"). Therefore,
for the birth-death system, the fluctuations are between a
configuration which acts as a "barrier", the birth configuration,
and a configuration acting as a "valley", the death configuration.
Note that a single-rate (namely, $a_f$=$b_b$) invariant
birth-death system is similar to the system studied by Doering and
Gadoua \cite{8}, where the derivatives of the two linear
potentials are sign opposite and equal in the absolute values.

Fig. 2 shows $F_{\gamma}(t)$ for a single-rate invariant
birth-death system and $n$=$10$. At short to intermediate times,
$F_{\gamma}(t)$ displays a peak that shifts towards larger times
with $\gamma$. This peak represents the exiting of the initial
population of configuration $A$, $\delta_{x,j}P_{A,0}$, for small
$\gamma$, and the overall initial condition for large $\gamma$. At
longer times and intermediate $\gamma$, a minimum in
$F_{\gamma}(t)$ appears as a function of $\gamma$ that represents
the fastest exit mainly of the initial $B$ population.
Accordingly, the minimum in the MFPT $\tau(\gamma)$ is a
consequence of the shape of $F_{\gamma}(t)$, and is therefore
reflected in higher moments of $F_{\gamma}(t)$ as well.
\begin{figure}[t]
\includegraphics[width=1.0\linewidth,angle=0]{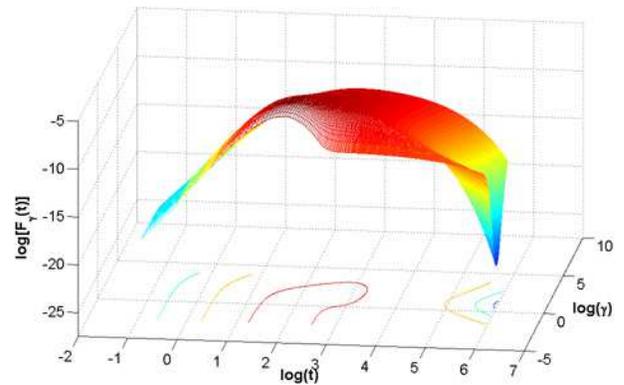}
\caption{$F_{\gamma}(t)$ as a function of $t$ and $\gamma$ on a
log-log-log plot (natural log), for a single rate invariant
birth-death system with $n$=$10$, and $a_f$=$b_b$=1. Also shown is
$F_{\gamma}(t)$ projections.}
\end{figure}

To study the RAP we start by computing the MFPT. In general, the
$s^{th}$ moment of $F(t)$ is obtained by inverting matrix
$\mathbf{H}$: ${\tau^s}=\int _0^\infty t^sF(t)dt=
s!\overrightarrow{U}_{2n}(-\mathbf{H})^{-s}\overrightarrow{P}_{2n}(0)$.
Using the projection operator techniques for $\mathbf{H}^{-1}$
blocks \cite{18}, $\tau$ reads
\begin{eqnarray}\label{2}
    \tau=\gamma\overrightarrow{U}_n(\mathbf{C_A}+\mathbf{C_B})\overrightarrow{P}_n(0)-~~~~~~~~~~~\nonumber\\
    -\overrightarrow{U}_n(\mathbf{C_A}\mathbf{A}+\mathbf{C_B}\mathbf{B})\overrightarrow{P}_n(0)/2,
\end{eqnarray}
where
$\mathbf{C_A}$=$[\mathbf{AB}-\gamma(\mathbf{A}+\mathbf{B})]^{-1}$,
$\mathbf{C_B}$=$[\mathbf{BA}-\gamma(\mathbf{A}+\mathbf{B})]^{-1}$,
and $[\overrightarrow{P}_n(0)]_j$=$\delta_{x,j}$. From Eq. (2) one
can calculate $\tau(\gamma)$ for the limiting cases
$\gamma\rightarrow 0$ and $\gamma\rightarrow \infty$. For
$\gamma\rightarrow 0$, $\tau$ is the average of the MFPT of the
uncoupled configurations, $A$ and $B$, $\tau$=$(\tau_A+\tau_B)/2$,
where
$\tau_M=-\overrightarrow{U}\mathbf{M}^{-1}\overrightarrow{P}_n(0)$
is the MFPT of configuration $M$. For $\gamma\rightarrow \infty$,
$\tau$ is the MFPT of an averaged fully coupled system, namely,
$\tau$=$-\overrightarrow{U}_n(\frac{\mathbf{A}+\mathbf{B}}{2})^{-1}\overrightarrow{P}_n(0)$.
These are the expected limiting behaviors of the MFPT
\cite{8}-\cite{12}, \cite{15,16}. RAP is expected for intermediate
flipping frequencies.

To obtain an analytical expression for the frequency that
minimizes $\tau(\gamma)$, $\gamma_{min}$, we search for an
extremum point (a minimum) of the function $\tau(\gamma)$, for an
invariant system and $n$=$2$. We find $\gamma_{min}$ to be a sum
of two terms:
\begin{equation}\label{3}
 \gamma_{min}=[\gamma_{min,1}\geq 0]+[\gamma_{min,2}\geq 0],
\end{equation}
where the notations on the right hand side of Eq. 3 mean that each
of the terms must be non negative to contribute to $\gamma_{min}$,
and
\begin{equation}\label{4}
    \gamma_{min,1}=\frac{a_f(3b_f^2-a_fb_b)-b_f(3a_f^2-b_fa_b)}{a_f(a_f-2b_b)-b_f(b_f-2a_b)},
\end{equation}
and
\begin{equation}\label{5}
    \gamma_{min,2}=\frac{a_f(b_f^2-a_fb_b)-b_f(a_f^2-b_fa_b)}{a_f(a_f+2b_b)-b_f(b_f+2a_b)}.
\end{equation}
We note that the smallest system that exhibits the RAP requires a
three site system, which is a specific case of the system shown in
Fig. 1, with, for example, $b_{1,2}=b_{2,1}\rightarrow\infty$.
However, in what follows we consider systems with finite
transition rates.

For the birth-death system Eq. (3) reduces to:
\begin{equation}\label{6}
 \gamma_{min}=\frac{a_f}{2-a_f/b_b}.
\end{equation}

The simple form of Eq. (6) provides an insight into the nature of
the RAP. It immediately implies the requirement $a_f/b_b<2$ for
RAP to occur. For $b_b\gg a_f$, $\gamma_{min}$=$(\tau_A)^{-1}$,
where $\tau_A$ is the first moment, $s$=$1$, of $F(t)$ for a death
system, $\tau_A^s=(n)_s/a_f^s$, where $(n)_s=(n+s-1)!/(n-1)!$.
This optimal frequency means that $\delta_{x,j}P_{A,0}$ has exited
the interval, on average, while the first configurational
transition occurred, and the same holds for $\delta_{x,j}P_{B,0}$,
for the second configurational transition. Because the probability
(particles) can exit the interval only when it is subject to the
$A$ configuration dynamics, a situation where the $A$
configuration is empty but not the $B$ configuration, means a
"waste" of time with regards to fastest interval exiting. This is
the case for $\gamma<\gamma_{min}$. For $\gamma>\gamma_{min}$, not
all $\delta_{x,j}P_{A,0}$ exited the interval, while the first
configurational transition occurred, meaning that another cycle of
flipping is required to exit the system. This leads again to a
"waste" of time with regards of fastest interval exiting. At
$\gamma=(\tau_A)^{-1}$, only one configurational change occurs,
and costs the minimal time for exiting the interval.

The special feature that $a_f/b_b<2$ is needed for RAP suggests
that the rate along the birth configuration must be, at least, as
fast as those along the death configuration for the RAP to be
obtained. For an invariant birth-death system to show the RAP, the
ratio $a_f/b_b$ must fulfil $a_f/b_b\leq 3$ asymptotically, which
is demonstrated in Fig. 3. Note that Fig. 3 spans both degrees of
freedom of the invariant birth-death case, the size $n$, and the
ratio $a_f/b_b$. Scaling the time, $\widetilde{t}=tb_b$, leads to
dimensionless rates $a_f/b_b$ and $\gamma/b_b$.
\begin{figure}[b]
\includegraphics[width=1.0\linewidth,angle=0]{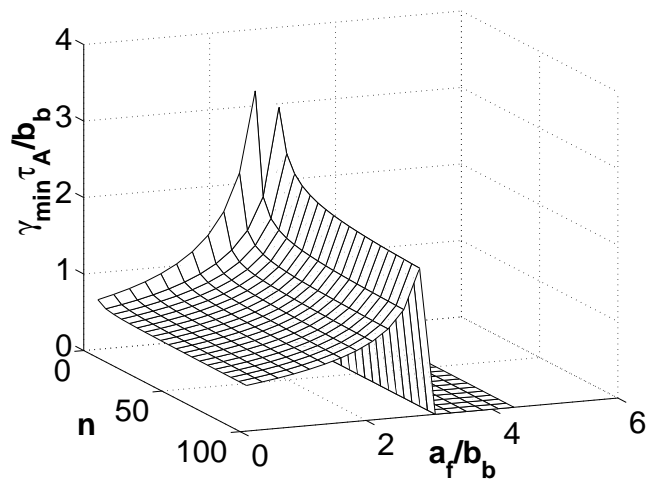}
\caption{$\gamma_{min}\tau_A/b_b$ as a function of $n$ and the
ratio $a_f/b_b$. When $a_f/b_b>3$ the RAP does not exist, for any
$n$.}
\end{figure}

We emphasize that even for the simple invariant birth-death system
that fulfills the demand that the $B$ configuration rate is much
larger than of configuration $A$ rates, the relation
$\gamma_{min}=(\tau_A)^{-1}$ might not be satisfied. To see that
we calculate $\gamma_{min}$ for a general birth-death system and
$n$=$2$ (see Fig. 2):
\begin{equation}\label{7}
 \gamma_{min}=\frac{\sqrt{a_{1,0}a_{2,1}}}{2-\sqrt{a_{1,0}a_{2,1}}/b_{1,2}},
\end{equation}
which for $\sqrt{a_{1,0}a_{2,1}}/b_{1,2}\rightarrow0$, reduces to
$\gamma_{min}=\sqrt{a_{1,0}a_{2,1}}/2$, where
$(\tau_A)^{-1}$=$a_{1,0}a_{2,1}/(a_{1,0}+a_{2,1})$. Moreover, when
$a_{1,0}\gg a_{2,1}$, $(\tau_A)^{-1}\approx a_{2,1}$,
$\gamma_{min}$ is unchanged, and is much larger than
$(\tau_A)^{-1}$, implying that more than one configurational
change occurs at the optimal flipping frequency. Reverse
substitution $\gamma$=$\gamma_{min}$ and $\gamma$=$(\tau_A)^{-1}$
into $\tau(\gamma)$, results in $\tau(\gamma=\gamma_{min})\approx
2/a_{2,1}$ and
$\frac{\tau(\gamma=\gamma_{min})}{\tau(\gamma=(\tau_A)^{-1})}\approx
0.8$. Note that for both values of $\gamma$, $\tau(\gamma)$ is
independent of $b_{1,2}$ to be compared with
$\tau(\gamma\rightarrow\infty)\approx
\frac{b_{1,2}}{a_{1,0}a_{2,1}}$. $\gamma_{min}$ has, therefore, a
general functional form that not necessarily coincides with the
MFPT of the faster configuration.

Going beyond the birth-death system, we first consider a case for
which $\mathbf{B}$=$\lambda \mathbf{A}$. From Eqs.(3)-(5) we have,
for $n$=$2$, $\gamma_{min}<0$ for any positive $\lambda$. From
numerical calculation we find that for $n>2$ there is no real
positive $\gamma_{min}$. Both the analytical and the numerical
results imply that a system for which $\mathbf{A}$ and
$\mathbf{B}$ commute does not exhibit the RAP.

The next case to be checked for the occurrence of the RAP is
obtained by setting, $a_f$=$b_b$=$k$. Using $\widetilde{t}=kt$,
the system transition rates are dimensionless, measured in units
of $k$. This procedure leads to the reduced Eqs.(4-5):
\begin{equation}\label{8}
    \gamma_{min,1}/k=\frac{v^2[3+u]-3v-1}
    {v[2u-v]-1},
\end{equation}
and
\begin{equation}\label{9}
    \gamma_{min,2}/k=\frac{v^2[1+u]-v-1}
    {3-v[2u+v]},
\end{equation}
where $u$=$a_b/k$ and $v$=$b_f/k$. Fig. 4a shows $\gamma_{min}/k$
as a function of $v$ for $u$=$1/2$. $\gamma_{min,1}/k$ displays a
maximum at $v_{max}$, which is easily recovered from Eq. (8). For
$v\leq v_{max}$, $\gamma_{min,1}/k$ increases, which reflects the
increase in the relative ability of the $B$ configuration to
"help" the fastest exiting of configuration $A$. On the other
hand, the decrease in $\gamma_{min,1}/k$ for $v\geq v_{max}$,
implies that the $B$ configuration movement towards the absorbing
end becomes fast enough "to stand on its own" for the
accomplishment of this task. A resonant-free zone occurs in the
region where $\tau_A\approx\tau_B$, and is followed by a short
resonant region, where both configurations are trap oriented,
namely, $u<1$ and
$v>1$. 
Fig. 4b shows that in the range $0\leq u<1$ $\gamma_{min,1}/k$ is
non-monotonic. Fig. 4c shows for $u$=$1$ the only non-zero
$\gamma_{min,1}/k>0$ which diverges as $\frac{1}{1-v}$ when
$v\rightarrow 1$, because for these system parameters
$\mathbf{B}$=$\mathbf{A}$. For $u>1$ and $v\leq 1$,
$\gamma_{min}/k$ has two asymptotic lines, which define a
resonant-free region. This happens when $\mathbf{B}\approx
\lambda\mathbf{A}$. For $u\gg v>1$, $\gamma_{min}/k\sim v/2$.
These two features are demonstrated in Fig. 4d.
\begin{figure}[t]
\includegraphics[width=1.0\linewidth,angle=0]{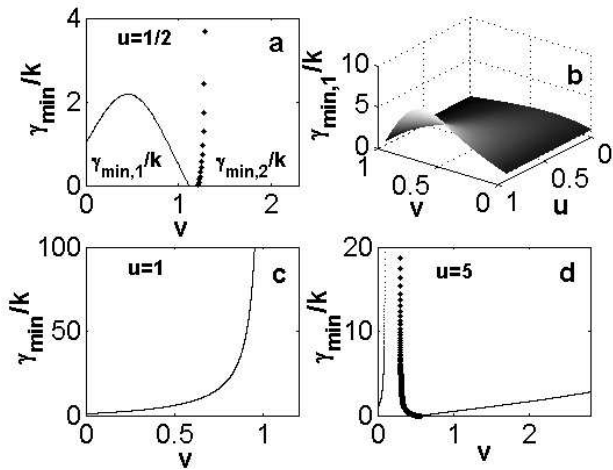}
\caption{{\bf a}: $\gamma_{min}/k$ for $a_f$=$b_b$=$k$, and
$u$=$1/2$, as a function of $v$. {\bf b}: $\gamma_{min,1}/k$ as a
function of $0\leq u,v<1$. {\bf c}: $\gamma_{min}/k$ for $u$=$1$
as a function of $v$. {\bf d}: $\gamma_{min}/k$ for $u$=$5$ as a
function of $v$.}
\end{figure}

The above analysis is of importance since the coupled ME with
$n$=$2$ can be used to model the kinetics of a conformationally
changing enzyme. Such extended Michaelis-Menten models are
appropriate for describing experiments performed on a single
molecule level [5]. If we assume that there are two enzyme
conformations, a specific stage of the enzymatic activity can be
described by Fig. 1. From Eqs. (8) and (9), and more generally
Eqs. (4) and (5), a relationship between the reaction rates and
the conformational flipping frequency can be established for an
optimal enzymatic activity. In addition, changes in the flipping
rate value near the resonant frequency, which can be achieved for
example by binding of other molecules to the enzyme, provide a
simple and efficient mechanism for regulating the enzymatic
activity, which is a well known issue in biology \cite{19}.
\begin{figure}[b]
\includegraphics[width=1.00\linewidth,angle=0]{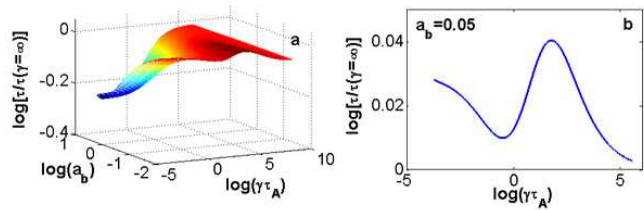}
\caption{{\bf a}: A log-log-log plot (natural log) of
$\tau(\gamma)/\tau(\gamma\rightarrow\infty)$ as a function of
$a_b$ and $\gamma\tau_A$ , for two absorbing ends in an invariant
system, with $a_f$=$3.5$, $b_f$=$0.175$, $b_b$=$1$, $n$=$7$, and
$x$=$4$. {\bf b}: Profile of the left figure for $a_b$=$0.25$
exposes a global minimum and a global maximum in
$\tau(\gamma)/\tau(\gamma\rightarrow\infty)$.}
\end{figure}

Finally, we study a system for which both ends are absorbing;
namely, the reflecting boundary is replaced by an absorbing one,
and the escape process starts at the middle site ($n$=$7$, and
$x$=$4$). The coupled invariant configurations are taken to have
an opposite bias, i. e. the transition rates of the $B$
configuration are $b_f$=$0.175$, $b_b$=$1$, which give rise to a
left side bias as defined in Fig. 1, whereas in the $A$
configuration $a_f$=$3.5$, and $a_b<a_f$ give rise to a right side
bias. A global minimum and a global maximum in
$\tau(\gamma)/\tau(\gamma\rightarrow\infty)$ can occur (Fig. 5b).
The minimum and maximum appear at the neighborhood of the points
$\gamma_{min}\tau_B=1$, and $\gamma_{max}\tau_A=1$, respectively.
This behavior is sensitive to the value of $a_b$ (Fig. 5a). For
$a_b\rightarrow 0$ the global extremum points reduce to local
extremum points. When $a_b\rightarrow a_f$,
$\tau(\gamma)/\tau(\gamma\rightarrow\infty)$ is a monotonically
increasing function of $\gamma\tau_A$.

We note that these boundary conditions for a fluctuating system
have been used to describe the translocation of a single stranded
DNA through a conformationally changing nanopore \cite{4}. For
this case no resonance occurred because of the physical conditions
that imposed the relation $\mathbf{B}$=$\lambda\mathbf{A}$.
However, for systems that are described by matrices $\mathbf{A}$
and $\mathbf{B}$ that do not commute and for two absorbing ends, a
change in $\gamma$ in the vicinity of the extremal points, leads
to a drastic change in the average time during which the system is
occupied, and, therefore, emphasizes the importance of the
frequency of fluctuation as a control parameter.

\begin{center}
{\bf {4. CONCLUSIONS}}\end{center}

To conclude, in this paper we revisited the resonant activation
phenomenon. We studied the origin of the RAP and the requirements
under which this phenomenon can be observed. We showed that for a
single rate invariant birth-death system the RAP is a consequence
of a general phenomenon, which is a non-monotonic behavior of
$F_{\gamma}(t)$ along the frequency axis for large time. We
characterized the conditions for which an invariant birth-death
system exhibits the RAP, and broaden these conditions by examining
more general systems. Relationship between the RAP and biological
activity was suggested. In addition, we introduced a new property
of the MFPT, the coexistence of a minimum and a maximum in the
flipping frequency dependent MFPT, $\tau(\gamma)$.


\end{document}